\documentclass[superscriptaddress,reprint,prd,amssymb,amsmath,nobibnotes,aps,showpacs,twocolumn]{revtex4-1}

\usepackage{graphicx}
\usepackage{epstopdf}
\usepackage{color}
\usepackage{subfig}
\usepackage{graphicx}
\usepackage{bm}
\usepackage{aas_macros}
\usepackage{appendix}
\usepackage{mathrsfs}

\def\be{\begin{equation}}
\def\ee{\end{equation}}
\def\bea{\begin{eqnarray}}
\def\eea{\end{eqnarray}}

\newcommand{\bicep}{B{\sc icep}1 }
\newcommand{\biceptwo}{B{\sc icep}2 }

\begin{document}

\title{Self-Calibration of BICEP1 Three-Year Data and Constraints on Astrophysical Polarization Rotation}

\author{J.P. Kaufman}
\affiliation{Department of Physics, University of California, San Diego, 9500 Gilman Drive, La Jolla, CA 92093-0424, USA}
\email[E-mail: ]{jkaufman@physics.ucsd.edu}
\author{N.J. Miller}
\affiliation{Observational Cosmology Laboratory, Code 665, Goddard Space Flight Center, 8800 Greenbelt Rd, Greenbelt, MD 20771, USA}
\author{M. Shimon}
\affiliation{School of Physics and Astronomy, Tel Aviv University, Tel Aviv 69978, Israel}
\affiliation{Department of Physics, University of California, San Diego, 9500 Gilman Drive, La Jolla, CA 92093-0424, USA}
\author{D. Barkats}
\affiliation{Joint ALMA Observatory, ESO, Santiago, Chile}
\author{C. Bischoff}
\author{I. Buder}
\affiliation{Harvard-Smithsonian Center for Astrophysics, 60 Garden Street MS 42, Cambridge, MA 02138, USA}
\author{B.G. Keating}
\affiliation{Department of Physics, University of California, San Diego, 9500 Gilman Drive, La Jolla, CA 92093-0424, USA}\author{J.M. Kovac}
\affiliation{Harvard-Smithsonian Center for Astrophysics, 60 Garden Street MS 42, Cambridge, MA 02138, USA}
\author{P.A.R. Ade}
\affiliation{Department of Physics and Astronomy, University of Wales, Cardiff, CF24 3YB, Wales, UK}
\author{R. Aikin}
\affiliation{Department of Physics, California Institute of Technology, Pasadena, CA 91125, USA}
\author{J.O. Battle}
\affiliation{Jet Propulsion Laboratory, Pasadena, CA 91109, USA}
\author{E.M. Bierman}
\affiliation{Department of Physics, University of California, San Diego, 9500 Gilman Drive, La Jolla, CA 92093-0424, USA}
\author{J. J. Bock}
\affiliation{Department of Physics, California Institute of Technology, Pasadena, CA 91125, USA}
\affiliation{Jet Propulsion Laboratory, Pasadena, CA 91109, USA}
\author{H.C. Chiang}
\affiliation{Astrophysics and Cosmology Research Unit, University of KwaZuluNatal, Durban, South Africa}
\author{C.D. Dowell}
\affiliation{Jet Propulsion Laboratory, Pasadena, CA 91109, USA}
\author{L. Duband}
\affiliation{SBT, Commissariat \`{a} lÕEnergie Atomique, Grenoble, France}
\author{J. Filippini}
\affiliation{Department of Physics, California Institute of Technology, Pasadena, CA 91125, USA}
\author{E.F. Hivon}
\affiliation{Institut dÕAstrophysique de Paris, Paris, France}
\author{W.L. Holzapfel}
\affiliation{Department of Physics, University of California at Berkeley, Berkeley, CA 94720, USA}
\author{V.V. Hristov}
\affiliation{Department of Physics, California Institute of Technology, Pasadena, CA 91125, USA}
\author{W.C. Jones}
\affiliation{Department of Physics, Princeton University, Princeton, NJ, 08544, USA}
\author{S.S. Kernasovskiy}
\affiliation{Stanford University, Palo Alto, CA 94305, USA}
\affiliation{Kavli Institute for Particle Astrophysics and Cosmology (KIPAC), Sand Hill Road 2575, Menlo Park, CA 94025, USA}
\author{C.L. Kuo}
\affiliation{Stanford University, Palo Alto, CA 94305, USA}
\affiliation{Kavli Institute for Particle Astrophysics and Cosmology (KIPAC), Sand Hill Road 2575, Menlo Park, CA 94025, USA}
\author{E.M. Leitch}
\affiliation{University of Chicago, Chicago, IL 60637, USA}
\author{P.V. Mason}
\affiliation{Department of Physics, California Institute of Technology, Pasadena, CA 91125, USA}
\author{T. Matsumura}
\affiliation{High Energy Accelerator Research Organization (KEK), Ibaraki, 305-0801, Japan}
\author{H.T. Nguyen}
\affiliation{Jet Propulsion Laboratory, Pasadena, CA 91109, USA}
\author{N. Ponthieu}
\affiliation{Institut dÕAstrophysique Spatiale, Universit\'{e} Paris-Sud, Orsay, France}
\author{C. Pryke}
\affiliation{Department of Physics, University of Minnesota, Minneapolis, MN, 55455, USA}
\author{S. Richter}
\affiliation{Department of Physics, California Institute of Technology, Pasadena, CA 91125, USA}
\author{G. Rocha}
\affiliation{Department of Physics, California Institute of Technology, Pasadena, CA 91125, USA}
\affiliation{Jet Propulsion Laboratory, Pasadena, CA 91109, USA}
\author{C. Sheehy}
\affiliation{University of Chicago, Chicago, IL 60637, USA}
\author{M. Su}
\affiliation{Department of Physics, Massachusetts Institute of Technology, 77 Massachusetts Avenue, Cambridge, MA, USA}
\affiliation{MIT-Kavli Center for Astrophysics and Space Research, 77 Massachusetts Avenue, Cambridge, MA, USA}
\author{Y.D. Takahashi}
\affiliation{Department of Physics, University of California at Berkeley, Berkeley, CA 94720, USA}
\author{J.E. Tolan}
\affiliation{Stanford University, Palo Alto, CA 94305, USA}
\affiliation{Kavli Institute for Particle Astrophysics and Cosmology (KIPAC), Sand Hill Road 2575, Menlo Park, CA 94025, USA}
\author{K.W. Yoon}
\affiliation{Stanford University, Palo Alto, CA 94305, USA}
\affiliation{Kavli Institute for Particle Astrophysics and Cosmology (KIPAC), Sand Hill Road 2575, Menlo Park, CA 94025, USA}



\date{\today}


\begin{abstract}

Cosmic Microwave Background (CMB) polarimeters aspire to measure the faint $B$-mode signature 
predicted to arise from inflationary gravitational waves.
They also have the potential to constrain cosmic birefringence,
rotation of the polarization of the CMB arising from parity-violating physics,
which would produce non-zero expectation values for the CMB's $TB$ and $EB$ spectra.  
However, instrumental systematic effects can also cause 
these $TB$ and $EB$ correlations to be non-zero.  
In particular, an overall miscalibration of the polarization orientation 
of the detectors produces $TB$ and $EB$ spectra which are degenerate with isotropic cosmological birefringence,
while also introducing a small but predictable bias on the $BB$ spectrum.
%
%
We find that \bicep three-year spectra, which use our standard calibration of
detector polarization angles from a dielectric sheet, 
are consistent with a polarization rotation of 
$\alpha = -2.77^\circ \pm 0.86^\circ \text{(statistical)} \pm 1.3^\circ \text{(systematic)}$.
We have revised the estimate of systematic error on the polarization rotation angle 
from the two-year analysis by comparing multiple calibration methods.  
We also account for the (negligible) impact of measured beam systematic effects.  
We investigate the polarization rotation for the \bicep 100 GHz and 150 GHz bands separately to 
investigate theoretical models that produce frequency-dependent cosmic birefringence. 
We find no evidence in the data supporting either these models or Faraday rotation of the CMB
polarization by the Milky Way galaxy's magnetic field. 
If we 
assume that there is no cosmic birefringence, we can use the 
$TB$ and $EB$ spectra to calibrate detector polarization orientations, thus reducing bias 
of the cosmological $B$-mode spectrum from leaked $E$-modes due to 
possible polarization orientation miscalibration.  
After applying this ``self-calibration" process, we find that the upper limit on the 
tensor-to-scalar ratio decreases slightly, from $r<0.70$ to $r<0.65$ at 95\% confidence.

\end{abstract}
\maketitle

\section{Introduction}
The cosmic microwave background (CMB) is a powerful cosmological probe; 
recombination physics, structure formation, and the cosmological
reionization history represent only a small subset of the phenomena 
probed by its temperature and polarization anisotropy. In addition, several aspects of fundamental physics can be
constrained by CMB observations, the most familiar of which are inflationary physics revealed via the imprint of 
primordial gravitational waves in the polarization of the CMB and the masses of neutrinos which can be probed via gravitational
lensing by dark matter.  These phenomena create $B$-mode polarization at the sub-$\mu K$ level.

Cosmological information can be extracted from the CMB's power spectra. Out of the six possible pairings of the temperature 
anisotropy $T$ and polarization $E$- and $B$-modes, only four have non-vanishing expectation values in the $\Lambda$CDM cosmological paradigm. 
The expectation values of the $TB$ and $EB$ cross-correlations vanish in the standard cosmological model 
due to parity symmetry but may assume non-vanishing values in the presence of  
systematics, astrophysical foregrounds, or, more interestingly, parity-violating 
departures from the standard models of electromagnetism and gravity. 
Any mechanism capable of converting $E$- to $B$-mode polarization 
necessarily leaks the $TE$ and $EE$ correlations to 
$TB$ and $EB$, respectively. 

A detection of $TB$ and $EB$ correlations of 
cosmological origin could undermine the fundamental assumptions of parity symmetry and Lorentz invariance by showing that our Universe possesses a small degree of chirality.
This phenomenon can be best revealed by CMB polarization where minuscule 
effects can accrue 
to observable levels over the $13.8$ Gyrs since CMB photons last 
scattered from the primordial plasma. This preferred 
chirality can be induced by the coupling of a pseudo-scalar field 
to either Chern-Simons-type terms in the electromagnetic 
interaction \cite{Carroll1990,Carroll1997,Carroll1998} 
or the Chern-Pontrayagin term in the case of gravitational interactions 
\cite{LueWangKamion1999,Alexander2005,Alexander2006}.  This work constrains the parameters in a scale-independent cosmological birefringence model as well as investigating frequency-dependent scale-independent models.  Current best constraints (not including this work) on scale-independent cosmological birefringence from CMB experiments are shown in Table \ref{table:CMB_CB}.  Though constraints on scale-dependent birefringence models have been reported with WMAP data \cite{YBSZ2009, Gluscevic2009, Gluscevic2010}, we do not provide such constraints in this work.  
A $3\sigma$ detection of cosmic birefringence was reported from combined 
WMAP, BOOMERanG, and \bicep two-year results (while explicitly excluding QUaD data) in \cite{Xia2010}.  
This work was later updated to include the impact of systematic effects at the levels reported by the three experiments and the significance reduced to $2.2\sigma$ \cite{Xia2012}.

\begin{table*}[t]
\caption{Previous rotation angle constraints from CMB experiments, following \cite{CMB_CB2012}.  Systematic uncertainties are shown in parentheses, where provided.}

\begin{tabular}{|c|c|c|c|}
\hline Experiment & Frequency (GHz) & $\ell$ range & $\alpha$ (degrees)\\
\hline WMAP7 \cite{Komatsu2011}& 41+61+94 & 2 - 800 & $-1.1 \pm 1.4 ~(\pm 1.5)$\\
\hline BOOM03  \cite{BOOM03}& 143 & 150 - 1000  & $-4.3 \pm 4.1$\\
\hline QUaD \cite{Wu2009}& 100 & 200 - 2000 & $-1.89 \pm 2.24 ~(\pm 0.5)$ \\
\hline QUaD \cite{Wu2009} & 150 & 200 - 2000 & $+0.83 \pm 0.94 ~(\pm 0.5)$\\
\hline
\end{tabular}

\label{table:CMB_CB}
\end{table*}


\bicep has set the most stringent constraints on the CMB's $B$-mode power spectrum \cite{Chiang2010, Barkats2013} in the multipole range $30 < \ell < 300$.  \bicep also measured the $TB$ and $EB$ power spectra in this range\cite{Chiang2010, Barkats2013}.  These $TB$ and $EB$ modes are extremely sensitive probes of departures from the standard cosmological model. 
In this work, we analyze the full \bicep three-year spectra \cite{Barkats2013} for evidence of polarization rotation,
considering systematic uncertainties including our primary and alternate polarization calibrations,
and exploring constraints on cosmological birefringence.  
We then use this polarization angle to ``self-calibrate" detector polarization orientations and calculate the tensor-to-scalar ratio from the ``self-calibrated" $BB$ spectrum.

The outline of the paper is as follows: Section \ref{s:CMB_spectra} contains a review of polarization rotation and how it affects the observed CMB power spectra.  The data sets and analysis procedure are described in Section \ref{s:data_analysis_method}.  Results and consistency checks are presented in Section \ref{s:RA_results}.
The impact of instrumental systematics is discussed in Section \ref{s:systematics}.  Consistency of the data with different birefringence models are in Section \ref{s:freq_dep}.  Application of ``self-calibration" and its effect on the tensor to scalar ratio, $r$, are in Section \ref{s:self_cal}, and we discuss our results in Section \ref{s:conclusion}.


\section{Polarization Rotation of the CMB Power Spectra} \label{s:CMB_spectra}

\begin{figure*}
\centering
\subfloat[$TB$]{\label{fig:rot_TB}\includegraphics[width=0.40\textwidth]{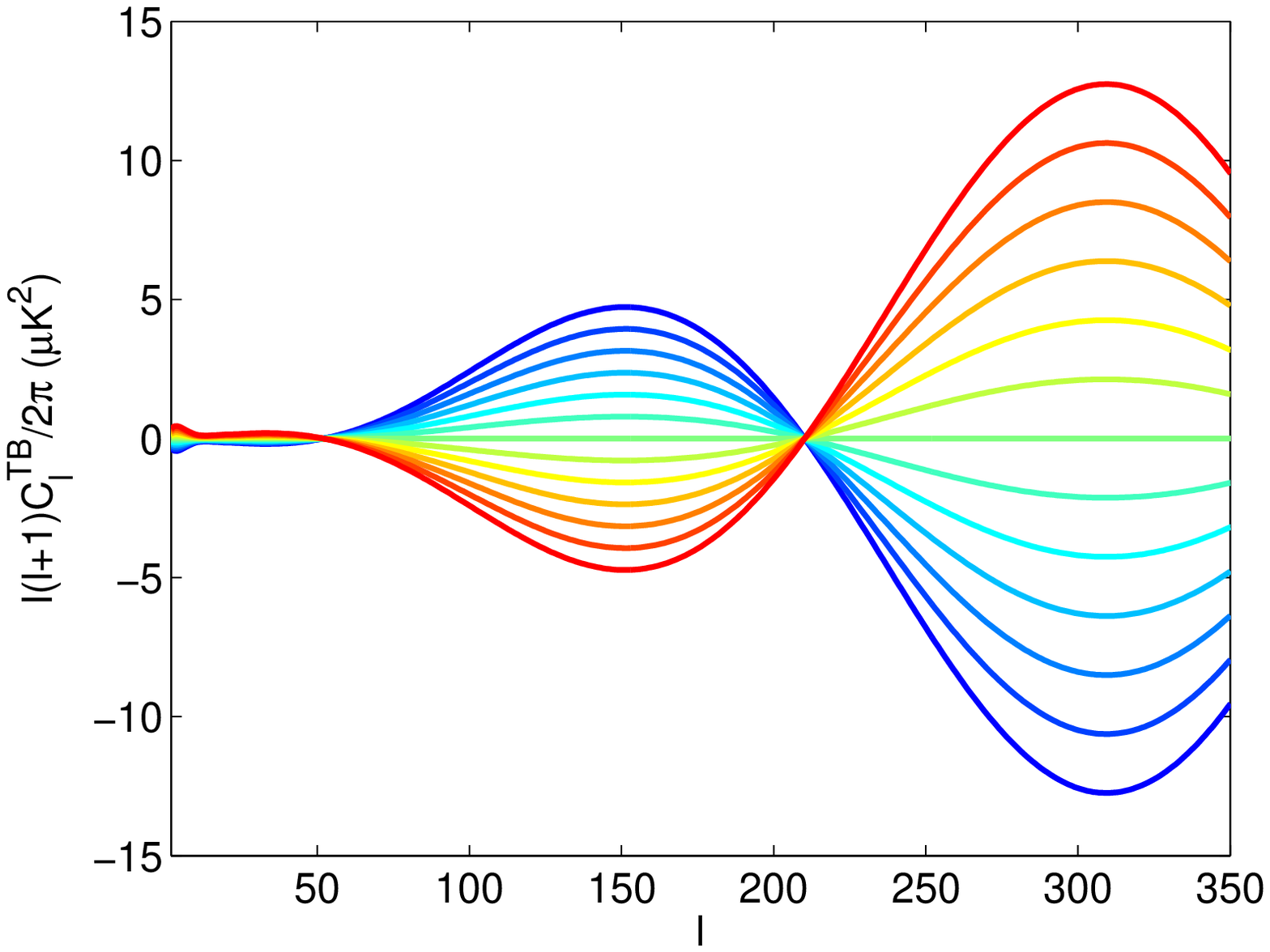}}
\subfloat[$EB$]{\label{fig:rot_EB}\includegraphics[width=0.40\textwidth]{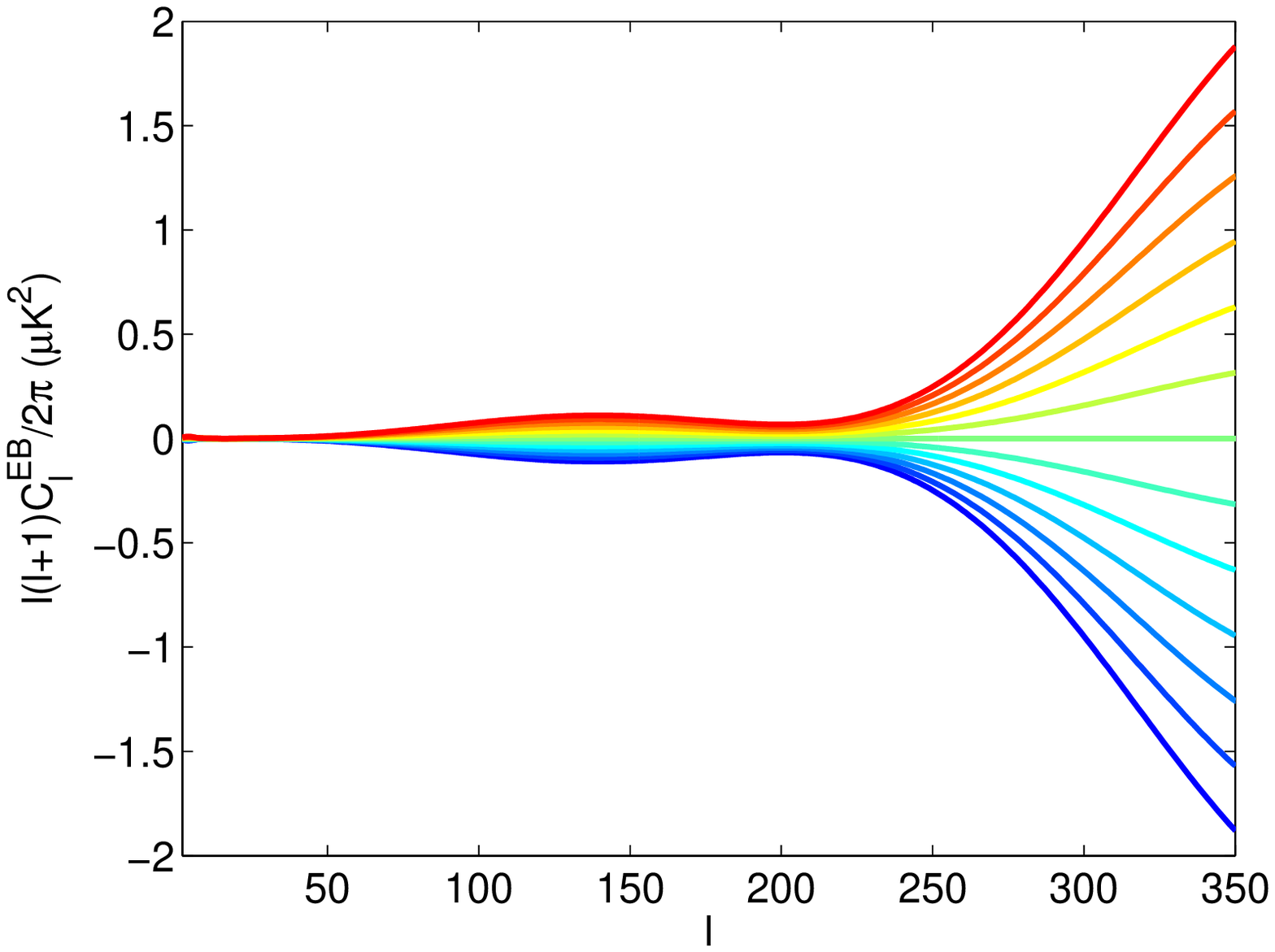}}
\caption{Standard $\Lambda$CDM power spectra after applying polarization rotation of -3$^\circ$ (blue) to +3$^\circ$ (red), in 0.5$^\circ$ steps, for $TB$ (left) and $EB$ (right).} 
\label{fig:theory_TBEB}
\end{figure*}

The CMB can be described by the statistical properties of its temperature and polarization.  $E$- and $B$-mode polarization can be formed from linear combinations of the Stokes $Q$ and $U$ parameters.
Maps of the temperature, $T$, and Stokes parameters, 
$Q$ and $U$, are expanded in scalar and spin $\pm 2$ spherical harmonics \cite{Seljak97,KKS1997} to obtain
\begin{eqnarray}
T(\hat{n}) &=& \sum_{\ell,m} a_{\ell m}^T Y_{\ell m} (\hat{n}) \nonumber \\
(Q \pm iU)(\hat{n}) &=& \sum_{\ell,m} a_{\pm 2, \ell m} \ _{\pm 2}Y_{\ell m} (\hat{n}),
\end{eqnarray}
where the $E$- and $B$-modes of polarization 
have expansion coefficients $a_{\ell m}^E$ and $a_{\ell m}^B$ which can be expressed 
in terms of the spin $\pm 2$ coefficients
\begin{equation}
a_{\pm 2, \ell m} = a_{\ell m}^E \pm i a_{\ell m}^B.
\end{equation}
The spherical harmonic coefficients, $a_{\ell m}$, are characterized by their 
statistical properties
\begin{eqnarray}
\left\langle a_{\ell m}^X \right\rangle &=& 0 \nonumber \\
\left\langle a_{\ell m}^{X*} a_{\ell' m'}^{X'} \right\rangle 
&=& C_{\ell}^{X X'} \delta_{\ell \ell'} \delta_{m m'},
\end{eqnarray}
where $X$ and $X'$ are either $T$, $E$ or $B$.  Here, $\langle a \rangle$ stands for the ensemble average.
The polarization modes, $E$- and $B$-, are pure parity states (even and odd, respectively) and thus 
the correlation over the full sky of the $B$-mode with either the temperature or 
$E$-mode polarization vanishes \cite{Seljak97,KKS1997}. However, if the polarization of the CMB is 
rotated, there will be a mixing between $E$ and $B$, subsequently inducing $TB$ and $EB$ power spectra (Figure \ref{fig:theory_TBEB}):
\begin{eqnarray}
C_{\ell}^{'TT} &=& C_{\ell}^{TT} \nonumber \\
C_{\ell}^{'TE} &=& C_{\ell}^{TE} \cos(2\alpha) \nonumber \\
C_{\ell}^{'EE} &=& C_{\ell}^{EE} \cos^2 (2\alpha) + C_{\ell}^{BB} \sin^2 (2\alpha) \nonumber \\
C_{\ell}^{'BB} &=& C_{\ell}^{EE} \sin^2 (2\alpha) + C_{\ell}^{BB} \cos^2 (2\alpha) \nonumber \\
C_{\ell}^{'TB} &=& C_{\ell}^{TE} \sin(2\alpha) \nonumber \\
C_{\ell}^{'EB} &=& \frac{1}{2}\left( C_{\ell}^{EE} - C_{\ell}^{BB} \right) \sin(4\alpha).
\label{eq:CBeqnsXia}
\end{eqnarray}
No assumption has been made here as to the source of this rotation, namely whether or not it is cosmological. 
In the literature, $\alpha$ is identified with the birefringence rotation angle (see \cite{Komatsu2011,Wu2009}), though here it is used to denote polarization rotation of any origin.

\section{Data and Analysis Methodology} \label{s:data_analysis_method}

\bicep observed for three years at the South Pole in three frequency bands: $100$, $150$ and $220$ GHz, and released two year results from 100 and 150 GHz frequency-combined spectra in \cite{Chiang2010} and three year frequency combined spectra in \cite{Barkats2013}.  Results from the \bicep 100, 150, and 220 GHz observations of the galactic plane are in \cite{Bierman2011} and from Faraday Rotation Modulators in \cite{Moyerman2013}.

We employ maximum-likelihood estimation for determining the best-fit polarization rotation angles of the power spectra following Equation \ref{eq:CBeqnsXia}.  We use two methods to construct the likelihoods, a Gaussian bandpower likelihood approximation and the Hamimeche-Lewis likelihood construction \cite{HL2008}.
 

\subsection{Data Sets}

We calculate rotation angles from the three-year frequency combined ``all-spectra" estimator, where ``all-spectra" is defined as $TE+EE+BB+TB+EB$.  We can break this down by frequency and by spectral estimator for consistency checks.  From this, we get four frequency subsets consisting of the two frequency auto-spectra: 100 GHz auto-spectra (denoted ``100") and 150 GHz auto-spectra (denoted ``150"), and the two frequency cross-spectra: 100 GHz cross-correlated with 150 GHz (denoted ``cross") and 150 GHz cross correlated with 100 GHz (denoted ``alt-cross").  Note that although the $EE$ and $BB$ spectra are identical for the ``cross" and ``alt-cross" data sets, the $TB$ and $EB$ spectra are not, e.g., $T^{100}B^{150} \neq T^{150}B^{100}$.

In addition, we have four spectral combinations to constrain $\alpha$: the $TB$ and $EB$ modes as well as the combination of $TB+EB$, and all-spectra: $TE$+$EE$+$BB$+$TB+EB$ since polarization rotation also affects $TE$, $EE$, and $BB$; however, from Equation \ref{eq:CBeqnsXia}, we can see that for small $\alpha$ the rotated $TE$, $EE$, and $BB$ deviate from the unrotated spectra by order $\alpha^2$ and thus their constraining power for $\alpha$ is much weaker than the $TB$ and $EB$ spectra, which are linear in $\alpha$.  
In addition, since they are quadratic in $\alpha$, the sign of $\alpha$ cannot be directly determined.  
$TB$ or $EB$ break this sign degeneracy.  
These are not independent estimators but are useful as any unexpected discrepancies can be used to test the validity of the analysis.

\begin{figure*}
\centering
\subfloat[$TB$]{\label{fig:real_TB}\includegraphics[width=0.80\textwidth]{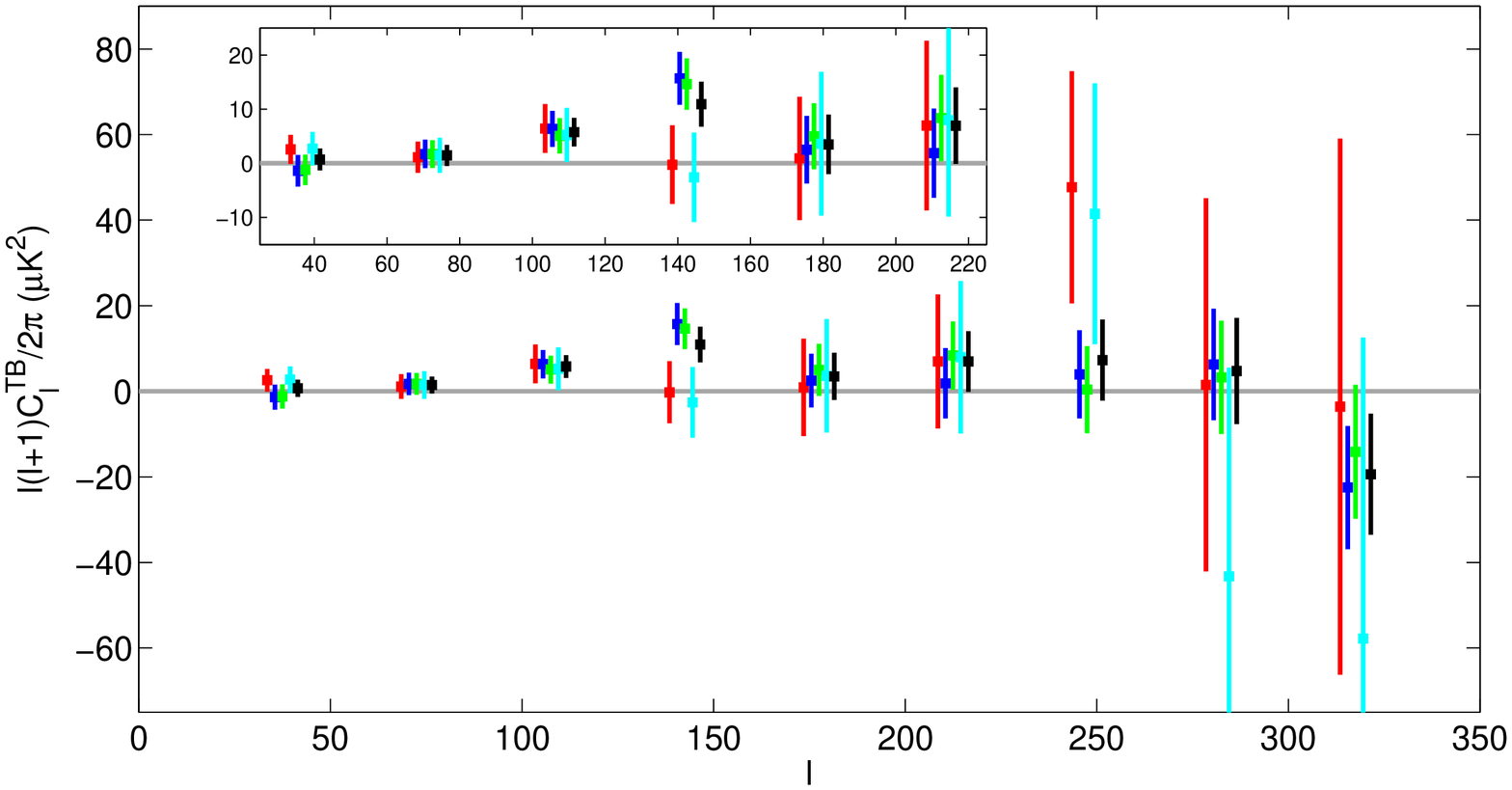}}\\
\subfloat[$EB$]{\label{fig:real_EB}\includegraphics[width=0.80\textwidth]{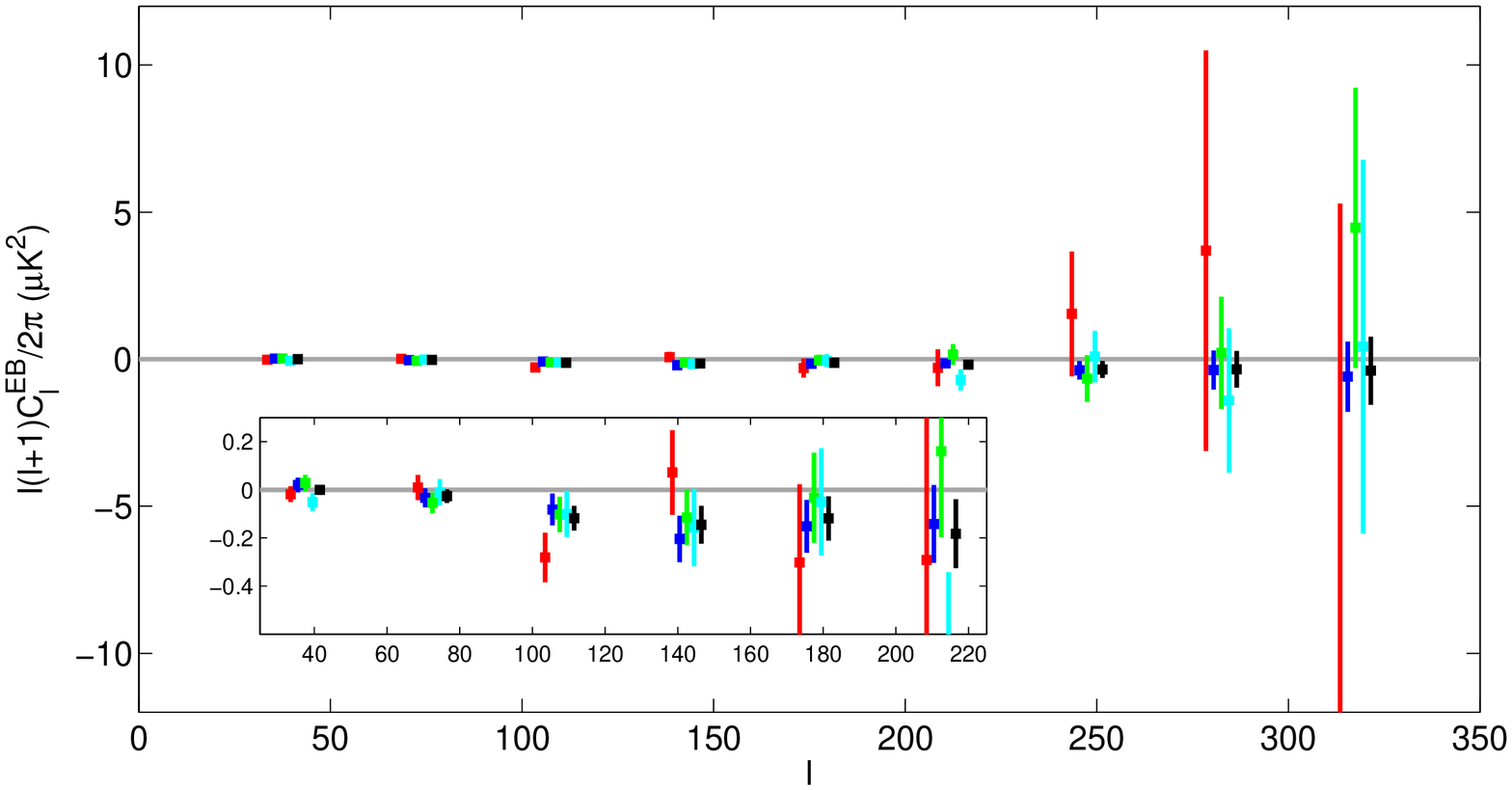}}
\caption{\bicep $TB$ and $EB$ power spectra for all frequency combinations: 100 GHz auto-spectra (red), 150 GHz auto-spectra (blue), 100 $\times$ 150 GHz cross-spectra (green), 150 $\times$ 100 GHz ``alt-cross" spectra (cyan), and frequency combined 100 $+$ 150 GHz spectra (black).  The points have been displaced in $\ell$ for clarity.} 
\label{fig:real_TBEB}
\end{figure*}


\subsection{Likelihood Analysis}

We employ two likelihood constructions for this analysis: a Gaussian bandpower likelihood approximation and the more accurate Hamimeche-Lewis (HL) likelihood approximation \cite{HL2008}.  The two likelihood constructions produce similar results, although we use the HL method for the final results since it more accurately treats cross-spectra covariances.  We test both likelihood constructions for any biases and, in simulations, we find they accurately recover known input rotation angles.

For both methods, we calculate $\chi^2 = -2\ln{\mathcal{L}}$, where $\chi^2$ is defined in Equation \ref{eq:chi2}, below.
We found the rotation angle that maximized the likelihood, and constructed 1$\sigma$ error bars by finding the minimum-width 68\% credible interval, assuming a uniform prior on $\alpha$, for both likelihood constructions.

\subsubsection{Gaussian Bandpower Likelihood Approximation} \label{s:standard_chi2}

This method was chosen due to its computational efficiency for isolating individual spectral estimators without including corresponding auto-spectra.  Here, the difference between the observed spectra and theory spectra including rotation 
\begin{eqnarray}
\Delta^{XY}_b(\alpha) &=& \hat{\mathcal{D}}_b^{XY} - \mathcal{D}_b^{XY}(\alpha),
\end{eqnarray}
is computed as a function of rotation angle, $\alpha$, for each \bicep multipole bin, where \bicep reports nine bins of uniform width $\Delta\ell =35$, with the first bin spanning $20 \le \ell \le 55$ and the ninth bin spanning $300 \le \ell <335$.
Here, $\hat{\mathcal{D}}_b^{XY}$ is the measured \bicep $XY$ power spectrum and $\mathcal{D}_b^{XY}(\alpha)$ is the theoretical rotated bandpower for $XY$ for a given $\alpha$.  We use $\mathcal{D}_b^{XY}$ to denote binned estimates of $\mathcal{D}_\ell^{XY} = \ell (\ell+1) C_\ell^{XY}/2\pi$.  Here, $XY = TB \text{ or } EB$ for each frequency combination.
The $\chi^2$ statistic is then constructed using
\begin{equation}\label{eq:chi2}
\chi^2_{XY}(\alpha) = \sum_{b b'} \Delta^{XY}_{b}(\alpha) \mathcal{M}^{-1}_{bb'} \Delta^{XY}_{b'}(\alpha),
\end{equation}
where $\mathcal{M}_{bb'}$ is the covariance between multipole bins $b$ and $b'$.  The covariance matrix, $\mathcal{M}$, was modeled as block-pentadiagonal, where only bandpowers separated by $\pm 2$ bins in $\ell$ are used for the calculation as the covariances between bandpowers with a larger separation are not well characterized (due to the finite number of simulations, 499 in total), but the contributions from these off-diagonal elements are known to be small.  



\subsubsection{Hamimeche-Lewis Method}

The Hamimeche-Lewis method is the bandpower likelihood approximation used in \cite{Barkats2013}.
  As before, the $\chi^2$ statistic is constructed as in Equation \ref{eq:chi2} but following the procedure outlined in  \cite{Barkats2013}.
One crucial difference between this method and the Gaussian bandpower likelihood approximation is that $XY$ includes all combinations of the spectra $X$ and $Y$.  For example, for $EB$, this method does not calculate the $\chi^2$ for $EB$ but actually the $\chi^2$ which includes $EB+EE+BB$ -- the comparison of the measured spectra to theoretical rotated spectra for $EB$, $EE$, and $BB$ simultaneously.  
To calculate the $\chi^2$ statistic for any ``pure" spectral combination using this method, we calculate the $\chi^2$ of the full spectral combination and subtract off the other spectral combinations.  For example, for $EB$:  $\chi^2_{EB} = \chi^2_{EE+BB+EB}-\chi^2_{EE}-\chi^2_{BB}$.

\section{Rotation Angle Results} \label{s:RA_results}

\begin{figure*}
\centering
\subfloat[$TB$]{\label{fig:comb_TB_w_alpha}\includegraphics[width=0.80\textwidth]{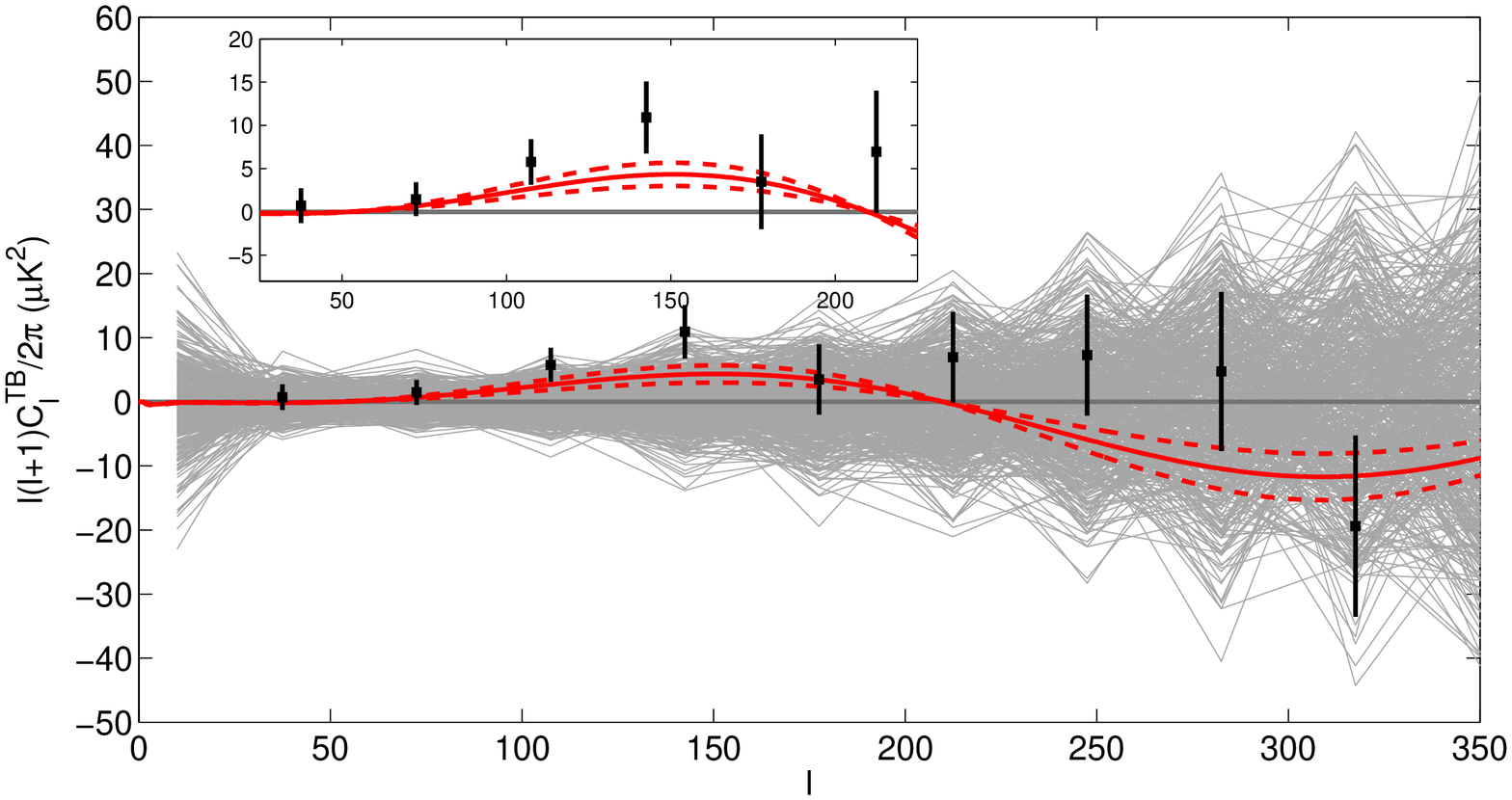}}\\
\subfloat[$EB$]{\label{fig:comb_EB_w_alpha}\includegraphics[width=0.80\textwidth]{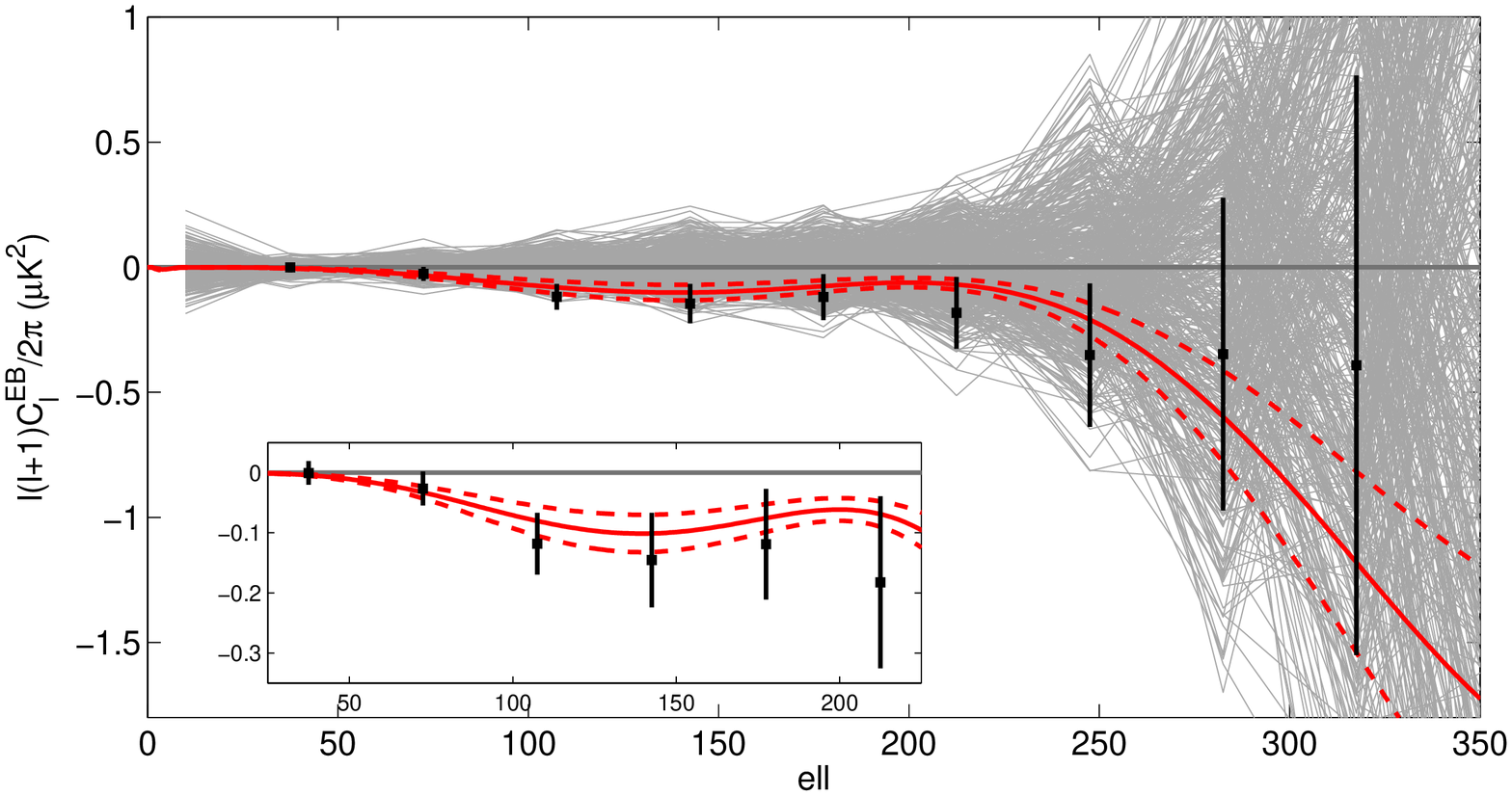}}
\caption{Frequency-combined three-year \bicep spectra (black points) shown with the theoretical rotated spectra from the best fit all-spectra rotation angle, $\alpha = -2.77^\circ \pm 0.86^\circ$ (red solid), the 1$\sigma$ confidence limits (red dashed), and the 499 simulation realizations (gray).  All simulations realization assume $\alpha=0$.} 
\label{fig:comb_TBEB_w_alpha}
\end{figure*}

The rotation angle, $\alpha$, was calculated using the HL method from the standard \bicep\ three-year
frequency combined spectra.
The best fit rotation angle is $\alpha = -2.77^\circ \pm 0.86^\circ$, where the quoted uncertainty is purely statistical.
These spectra use our standard calibration of detector polarization angles from a dielectric sheet; 
systematic uncertainty on this calibration is discussed below in
Section \ref{s:systematics}.
\begin{figure}[h]
\includegraphics[width=0.40\textwidth]{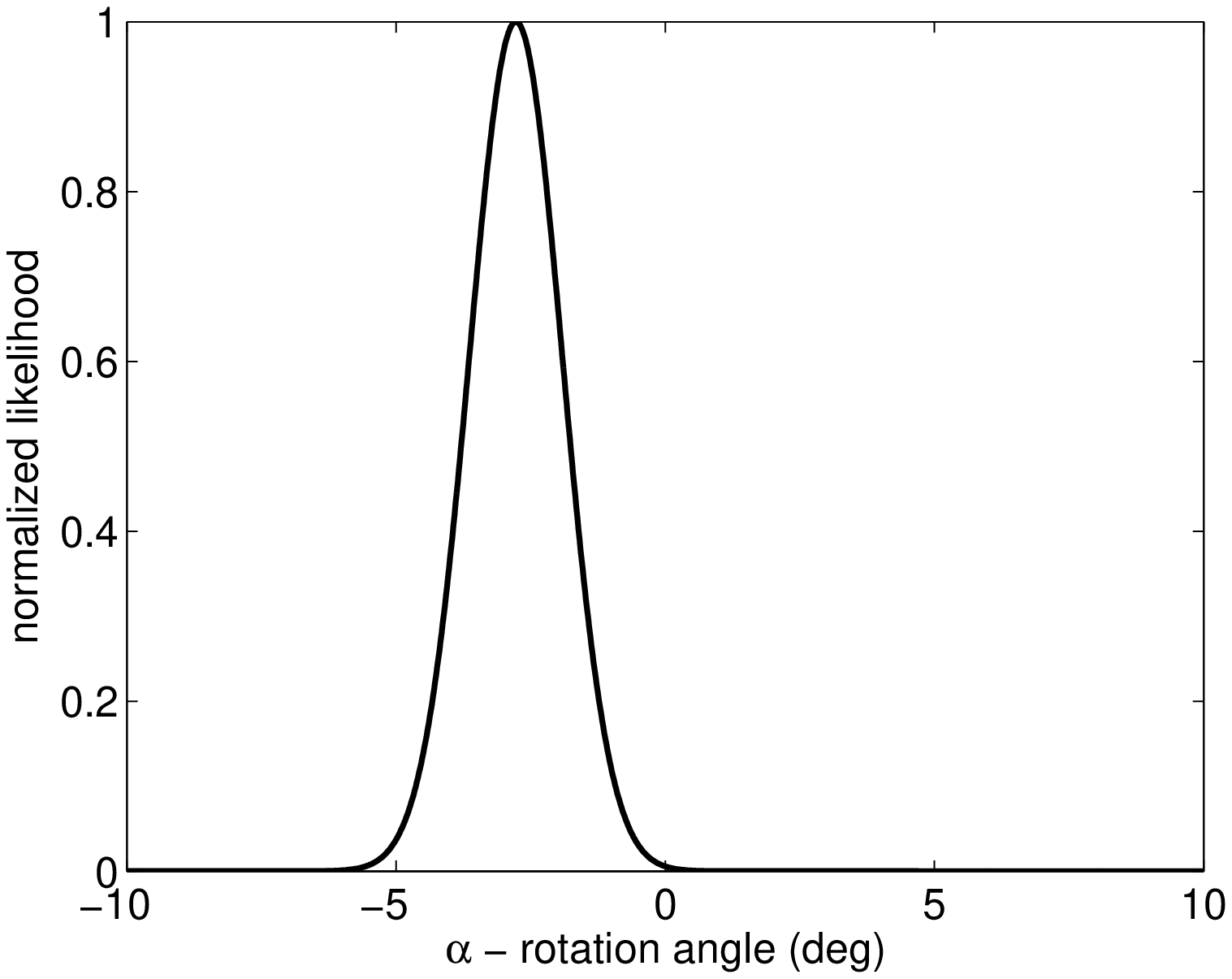}
\caption{The peak-normalized Hamimeche-Lewis likelihood for the all-spectra (``$TEB$") rotation angle.  The maximum likelihood value is -2.77$^\circ$ and the 68\% confidence limits are $\pm 0.86^\circ$ from the peak value, corresponding to 3.22$\sigma$ statistical significance.}
\label{fig:like_comb_TEB}
\end{figure}
Figure \ref{fig:like_comb_TEB} plots the peak-normalized HL likelihood and Figure \ref{fig:comb_TBEB_w_alpha} shows the best fit rotation angle spectra plotted compared to the \bicep three-year data and 499 simulated $\Lambda$CDM realizations (i.e., with $\alpha =  0$).

\subsection{Consistency Between Analysis Methods}

\begin{figure}
\includegraphics[width=0.40\textwidth]{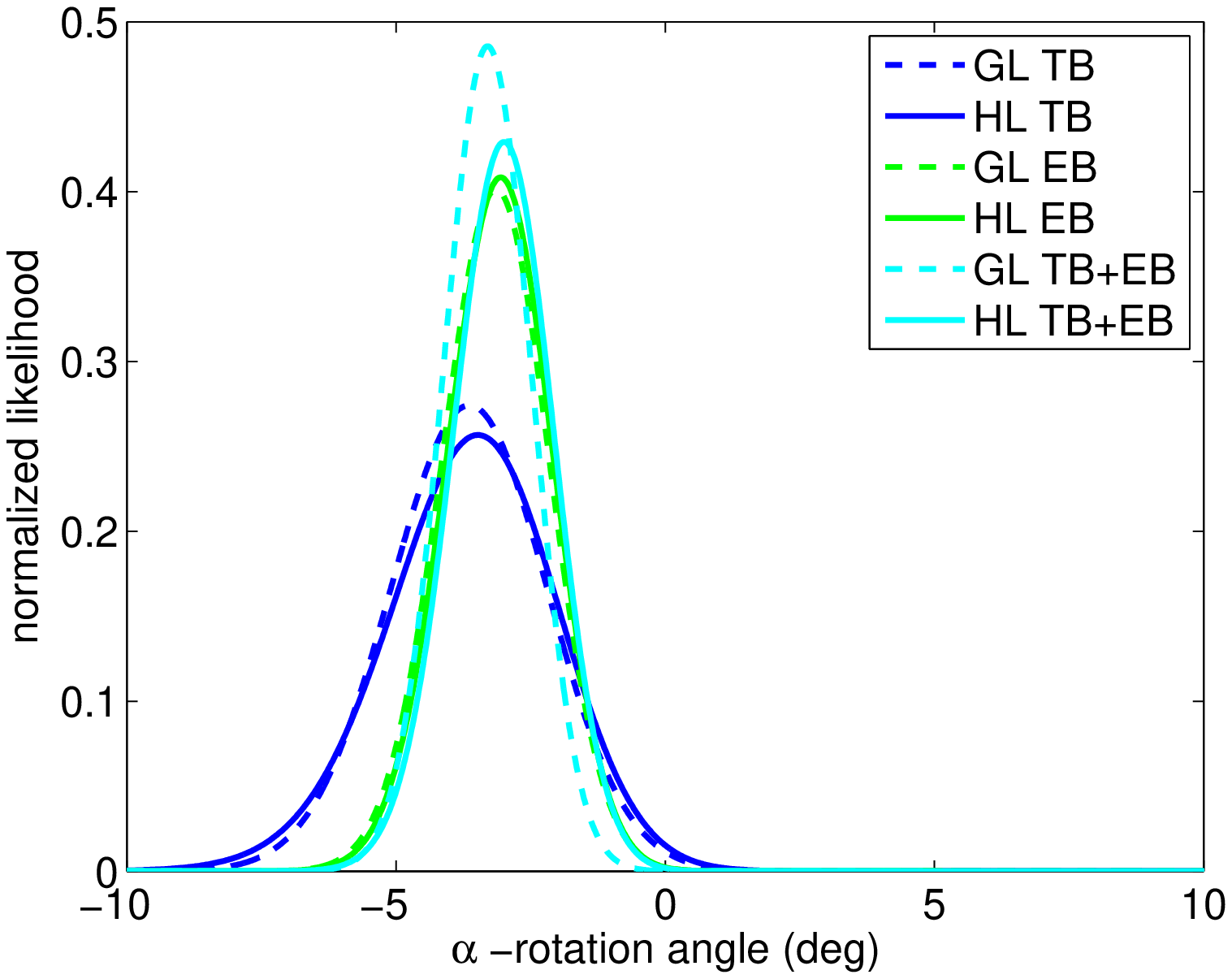}
\caption{Comparison of the two likelihood methods employed for the rotation angle calculation for the frequency combined data.  Likelihoods computed from the Gaussian bandpower likelihood approximation are shown as dashed lines and the Hamimeche-Lewis method likelihoods are shown as solid lines.  The $TB$ likelihood is in blue, $EB$ is in green, and the $TB+EB$ likelihood is cyan.  The likelihoods have been normalized by their integral over $\alpha$.} 
\label{fig:comp_like_method}
\end{figure}

To check that the rotation angle is not dependent on the analysis method, polarization rotation angles derived from the two analysis methods were compared.  In Figure \ref{fig:comp_like_method}, the likelihoods calculated for $TB$, $EB$, and $TB+EB$ for the frequency combined spectra for both the HL and Gaussian bandpower likelihood approximations are overplotted.  For all three available spectral estimators, the analysis methods agree to within $0.32\sigma$, $0.30\sigma$, and $0.18\sigma$ for $TB$, $EB$, and $TB+EB$, respectively.

\subsection{Consistency Between Frequencies}

\begin{figure}
\includegraphics[width=0.40\textwidth]{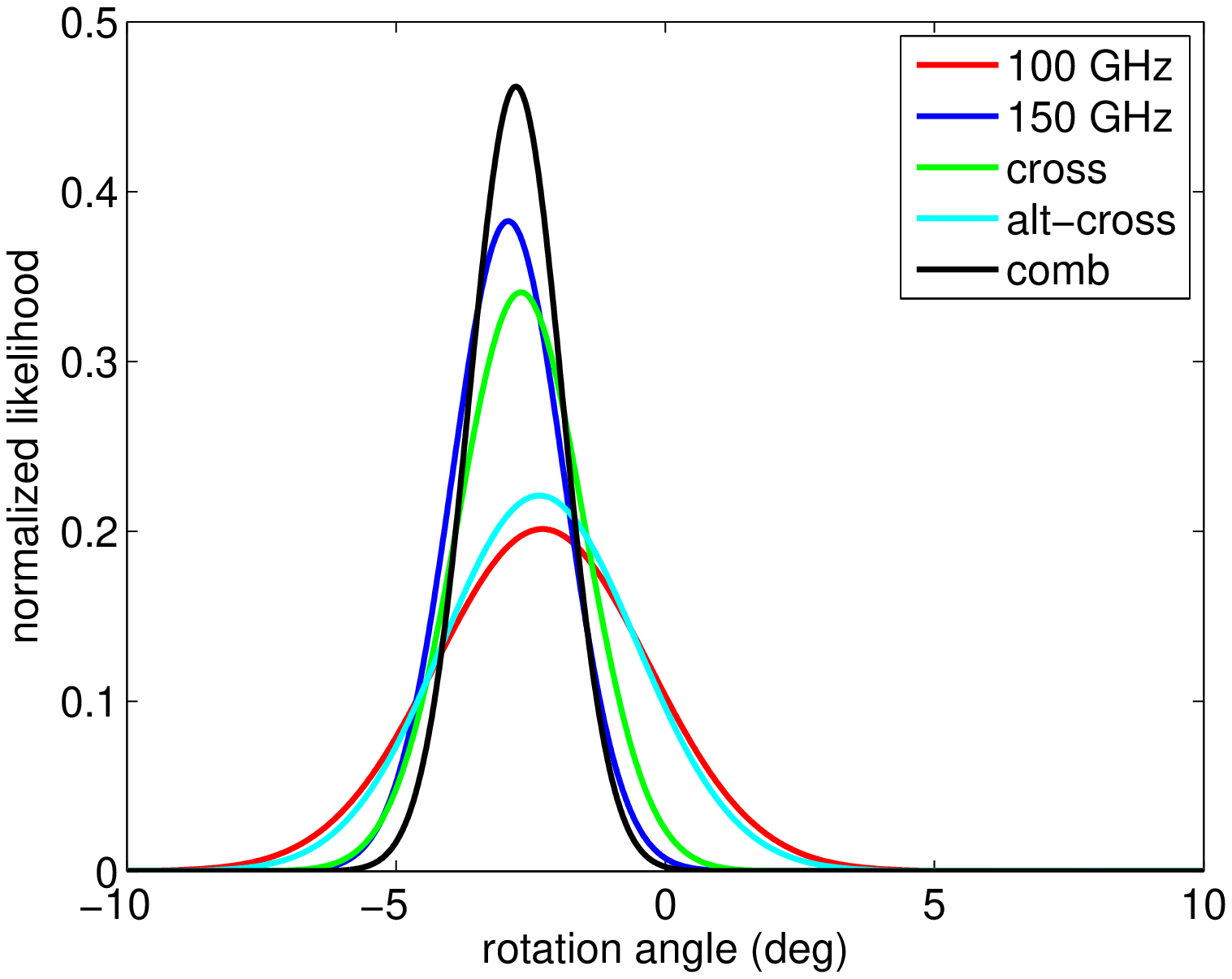}
\caption{Comparison of integral-normalized Hamimeche-Lewis method likelihoods of all-spectra (TEB) rotation angles for 100 GHz auto-spectra (red), 150 GHz auto-spectra (blue), 100 $\times$ 150 GHz cross-spectra (green), 150 $\times$ 100 GHz ``alt-cross" spectra (cyan), and frequency combined spectra (black).} 
\label{fig:comp_alpha}
\end{figure}

For consistency, the different frequency combinations were checked to determine if they have similar rotation angles.  Table \ref{table:BicepCB} shows the calculated rotation angles from each frequency data set and for all four spectral estimators.  Figure \ref{fig:comp_alpha} shows the HL likelihoods for the all-spectra (``$TEB$") rotation angles for each data set.

\begin{table}[h]
\caption{Maximum likelihood value and $1\sigma$ error for $\alpha$.  All numbers are in degrees.}
\begin{tabular}{|c|c|c|c|c|}
\hline Dataset & $TB$ only & $EB$ only & $TB+EB$ & all spectra \\
\hline $100$ GHz & $-1.79^{+3.18}_{-3.14}$ & $-3.53^{+2.38}_{-2.26}$ & $-2.27^{+2.06}_{-1.98}$ & $-2.27^{+2.06}_{-2.02}$\\
\hline $150$ GHz & $-4.37^{+1.92}_{-1.78}$ & $-2.95^{+1.20}_{-1.18}$ & $-3.13^{+1.14}_{-1.12}$ & $-2.91^{+1.06}_{-1.04}$\\
\hline cross & $-3.93^{+1.84}_{-1.74}$ & $-2.55^{+1.68}_{-1.60}$ & $-2.83^{+1.28}_{-1.24}$ & $-2.67^{+1.20}_{-1.18}$\\
\hline alt-cross & $-2.71^{+3.52}_{-3.74}$ & $-3.25^{+2.26}_{-2.20}$ & $-3.45^{+2.24}_{-2.18}$ & $-3.15^{+1.96}_{-2.00}$\\
\hline comb & $-3.47^{+1.66}_{-1.56}$ & $-3.05^{+1.00}_{-0.96}$ & $-2.99^{+0.94}_{-0.92}$ & $\mathbf{-2.77^{+0.86}_{-0.86}}$\\
\hline
\end{tabular}
\label{table:BicepCB}
\end{table}

\subsection{Consistency with Planck Temperature Maps}

The $TB$ spectrum estimate of the polarization rotation explicitly depends on the \bicep measurement of temperature.  To check for systematics in the $TB$ power spectrum, we replace \bicep maps with Planck temperature maps \cite{Planck2013} for both 100 and 150 GHz and find the recovered angles agree to within $0.2\sigma$.


\section{Impact of Instrumental Systematic Effects} \label{s:systematics}


\bicep was the first experiment designed specifically to measure the $B$-mode power spectrum in order to constrain the inflationary cosmological model \cite{Keating2003}.  
Accordingly, the analysis of instrumental systematics focused on potential bias of the $BB$ spectrum and the tensor-to-scalar ratio with a benchmark of $r=0.1$ \cite{Takahashi2010}. 
Here, we extend the analysis to include the impact of measured systematics on the $TB$ and $EB$ power spectra for the three-year data set.

\subsection{Polarization Angle Calibration}
An error in the detector polarization angles used for map-making is the only systematic which is completely degenerate with a rotation due to isotropic cosmic birefringence, and the only systematic capable of producing self-consistent $TB$ and $EB$ power spectra \cite{Yadav2013}.
This calibration requirement is much more stringent when attempting to measure $\alpha$ than for $r$.

%
%


Calibrating detector angles for CMB polarimeters is very challenging.
Some commonly employed methods include man-made calibrators, such as a polarizing dielectric sheets \cite{Takahashi2010,Keating2003} or polarization-selecting wire grids \cite{Tajima2012,George2012}, and observations of polarized astronomical sources \cite{Farese2004,QUIET2013}.
Man-made polarization calibration sources suffer from a host of challenges: they are often situated in the near-field of the telescope, they can be unstable over long timescales, and they can be cumbersome to implement and align.
Astronomical sources are not visible from all observatories and even the best characterized sources have orientations measured to an accuracy of only $0.5^\circ$ \cite{Aumont2009}.
In addition, the brightness of both astronomical and man-made calibration sources can overload the detectors, forcing them into a non-linear response regime \cite{KSY}.

\bicep employed several hardware calibrators to measure detector polarization angles.
The primary calibration comes from a dielectric sheet calibrator (DSC), described in detail in \cite{Takahashi2010}, but additional calibrations were made using sources with polarizing wire grids in the near and far field.
The \bicep beam size and observatory location prevented polarization calibration using astronomical sources.

The polarization angle measurement from the DSC was performed the most frequently and is the best studied, which is why it was chosen for results in \cite{Chiang2010,Barkats2013}, as well as this work.
Repeated measurements during each observing season produced polarization angles that agree with an rms error of $0.1^\circ$.
However, measurements taken before and after focal plane servicing between the 2006 and 2007 observing seasons show an unexplained rotation of $1^\circ$ in the polarization angles.
There is also some uncertainty in translating the results of the DSC measurement to parameters appropriate for CMB analysis. 
The details of the polarized signal from the dielectric sheet depend on the near field response of each detector, which is not well characterized.


We also consider two alternate calibrations for the detector polarization angles, which were both described in \cite{Takahashi2010} as methods to measure the cross-polar response of the detectors.
The first is a modulated broadband noise source, broadcasting via a rectangular feedhorn located behind a polarizing wire grid. 
The source is located on a mast at a range of 200 meters.
We measure the detector response as a function of angle by scanning over the source with 18 different detector orientations.
The advantage of this method is that the source is in the far field of the telescope.
A challenge is that the observations require the use of a flat mirror, complicating the pointing model.
In addition, it takes a significant amount of time to perform scans at all 18 orientations, which makes it more difficult to maintain stable source 
brightness.
For \biceptwo\!\!, we have invested significant effort in improving polarization orientation calibrations
with the far-field broadband noise source, both by developing a high-precision rotating polarized source and
improving the pointing model used for calibration analysis.  These improvements were motivated by \bicep\
experience with systematic uncertainties on both the DSC and broadband source calibrations.

Another polarization angle calibrator consists of a wire grid covering a small aperture that is chopped between an ambient temperature absorber and cold sky.
For calibrations, this source is installed in the near field of the telescope and the wire grid is rotated to measure detector polarization angles.
The interpretation of results from this source has significant uncertainty because the small aperture probes only a small fraction of the detector near field response yet the results are extrapolated to the full beam response.

Table \ref{table:CB_cals} lists the values of $\alpha$ measured from maps made using each of the polarization angle calibrations.
Also shown is the result obtained if we simply assume that the detector polarization angles are as-designed.
These derived $\alpha$ values are qualitatively consistent with the average difference in the detector polarization angles between any two calibration methods, though the details depend on how each detector is weighted in the three-year maps.
Besides the global rotations between each calibration method, which contributes to the variation in $\alpha$, the per-detector polarization angles show scatter of $0.6$--$0.9^\circ$ between methods, much larger than the $0.1^\circ$ consistency seen from repeated measurements using the DSC.
Despite this scatter, we can observe significant structure in the pattern of polarization angles from detector to detector, which is not present in the as-designed angles.

From consideration of the $1.14^\circ$ difference between alpha as derived from the DSC calibration and the mean of the three alternate calibrations, which have very different sources of systematic uncertainty, as well as the $1^\circ$ shift observed in the DSC calibration results between observing seasons, we assign a calibration uncertainty of $1.3^\circ$ on the overall orientation from the DSC calibration.  We believe this upward revision of the $0.7^\circ$ uncertainty quoted for this same calibration in \cite{Takahashi2010} is warranted by the tension with the alternate calibrations.
While this systematic error is larger than the $0.86^\circ$ statistical error on $\alpha$, we stress the fact that the polarization angle calibration is quite a bit better than what is needed to meet the $r=0.1$ benchmark for the primary \bicep science goal.

\begin{table}[h]
\caption{Polarization rotation angles derived using different detector polarization angle calibrations: the dielectric sheet calibrator (DSC), the far-field wire grid broadband noise source, the near-field wire grid aperture source, and assuming the polarization angles are as-designed. 
}

\begin{tabular}{|c|c|c|}
\hline Calibration method & near/far-field & $\alpha$ (degrees)\\
\hline DSC & near & -2.77 \\
\hline wire grid broadband source & far & -1.71 \\
\hline wire grid aperture source & near & -1.91 \\
\hline as-designed & --- & -1.27 \\
\hline
\end{tabular}

\label{table:CB_cals}
\end{table}

In Section \ref{s:self_cal}, we adopt a different approach and ``self-calibrate'' the polarization orientations by rotating the polarization maps to minimize $\alpha$.
Note that the calibration uncertainty on $\alpha$ applies only when we use the DSC calibrated maps and attempt to measure astrophysical polarization rotation.
To judge how well the self-calibration procedure can debias the $B$-mode map, only the statistical error is relevant.

\subsection{Differential Beam Effects}

Differential beam mismatches potentially mix $E$-modes and $B$-modes or leak intensity to either $E$- or $B$-mode polarization.  Here, we investigate the impact of differential beam size, differential relative gain, differential pointing, and differential ellipticity on the derived rotation angle.

Beam systematics affect the $EB$ spectra 
in a different way than $TB$ spectra \cite{Shimon2008,Miller2009}.  As a result, the scale 
dependence of the beam systematic polarization will imply a different {\it effective} 
rotation angle in the $TB$ spectrum versus the $EB$ spectrum, for a fixed $\ell-$range.

The \bicep beams were measured in the lab prior to deployment using a source in the far-field (50 meters from the aperture) and each observing season during summer calibration testing. 
Beam maps were fit to a two dimensional elliptical gaussian model which included a 
beam location, width, ellipticity, and orientation of the major axis of 
the ellipse with respect to the polarization axes.

\subsubsection{Differential Beam Size}
Though the differential beam size effect can leak temperature to polarization, due to circular symmetry it will not break the parity of the underlying sky and thus cannot generate the parity-odd $TB$ and $EB$ modes \cite{Shimon2008}.

\subsubsection{Differential Relative Gain}

As with differential beam size, circular symmetry is preserved by differential relative gain and thus there is no breaking of the parity of the sky which would generate $TB$ and $EB$ modes \cite{Shimon2008}.  We ran differential gain simulations using observed values and random values drawn from a gaussian distribution with an rms of 1\%.  None of the simulations showed polarization rotation greater in magnitude than 0.25$^\circ$.

A significant difference between the \bicep two-year results reported in \cite{Chiang2010} and the \bicep three-year power spectra is that the three-year spectra undergo relative gain deprojection which reduces $B$-mode contamination due to this systematic to negligible levels \cite{Barkats2013}.

\subsubsection{Differential Pointing}
The effect of differential pointing is analytically calculated using the measured 
magnitude and direction of beam offsets with the expected amount of false 
$BB$ power scaling as the square of the magnitude of the differential pointing, following the construction in  \cite{Shimon2008}. 
The upper limit on differential pointing error was estimated to be $<1.3\%$ of the beam size. 
While this was found to be the dominant beam systematic effect for the \bicep 
limit on $r$ \cite{Barkats2013}, it is clear from Figure \ref{fig:tbeb_ellip} that differential pointing does not induce $TB$ or $EB$, and has a negligible effect on the polarization rotation angle estimation.  This was calculated for the ``worst-case" scanning strategy and therefore provides very conservative bounds on the $TB$ and $EB$ produced. 


\subsubsection{Differential Ellipticity}

Differential ellipticity values were derived by fitting each measured beam in a detector pair for ellipticity and then differencing those values for the two detectors in a pair. The fits were generally not repeatable when the telescope was rotated about its boresight angle, so only upper limits on differential ellipticity are quoted.  The \bicep estimated differential ellipticity is $<0.2\%$.

\begin{figure*}[t]
\centering
\subfloat[$TB$]{\label{fig:tb_ellip}\includegraphics[width=0.40\textwidth]{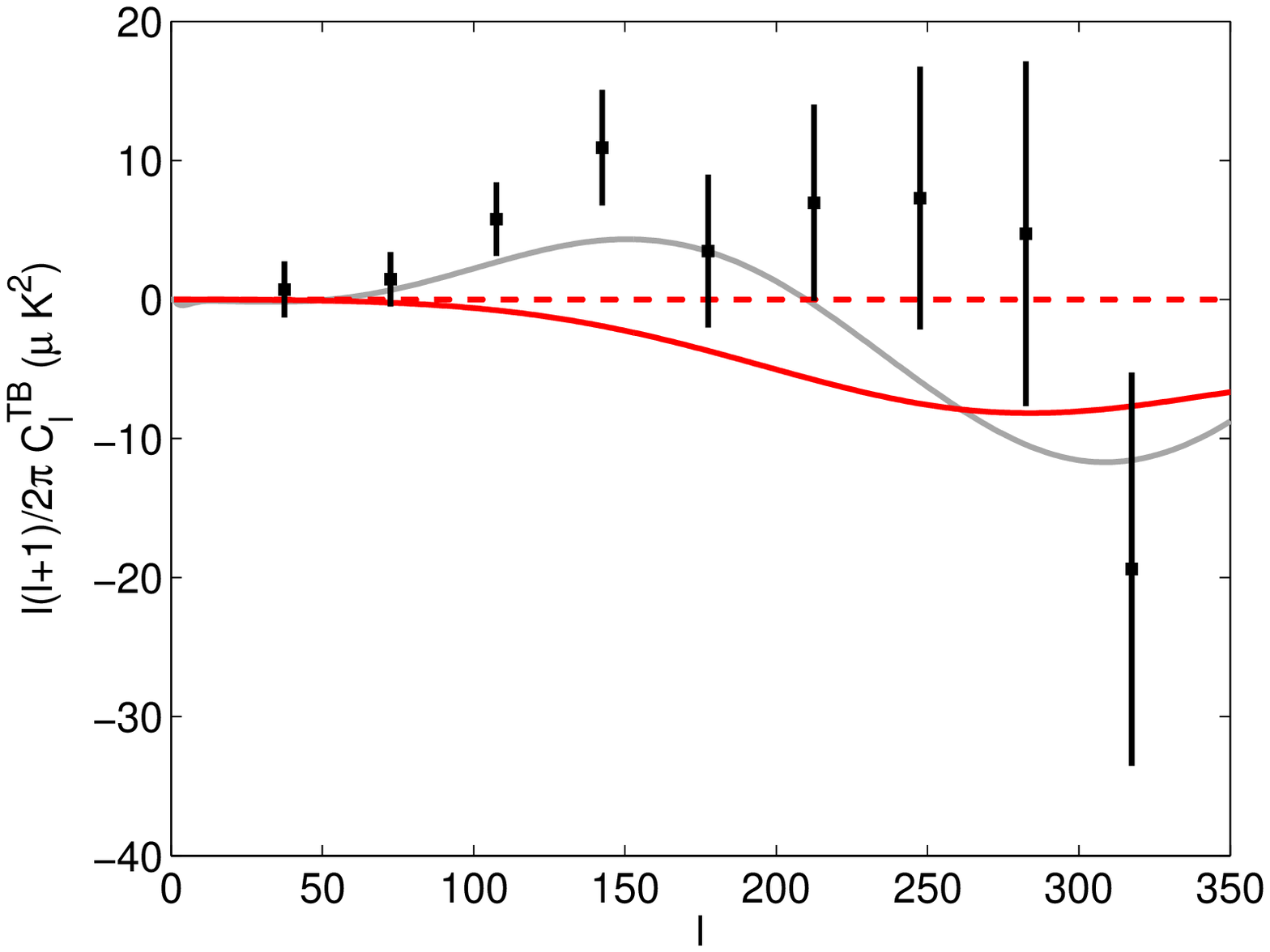}}
\subfloat[$EB$]{\label{fig:eb_ellip}\includegraphics[width=0.40\textwidth]{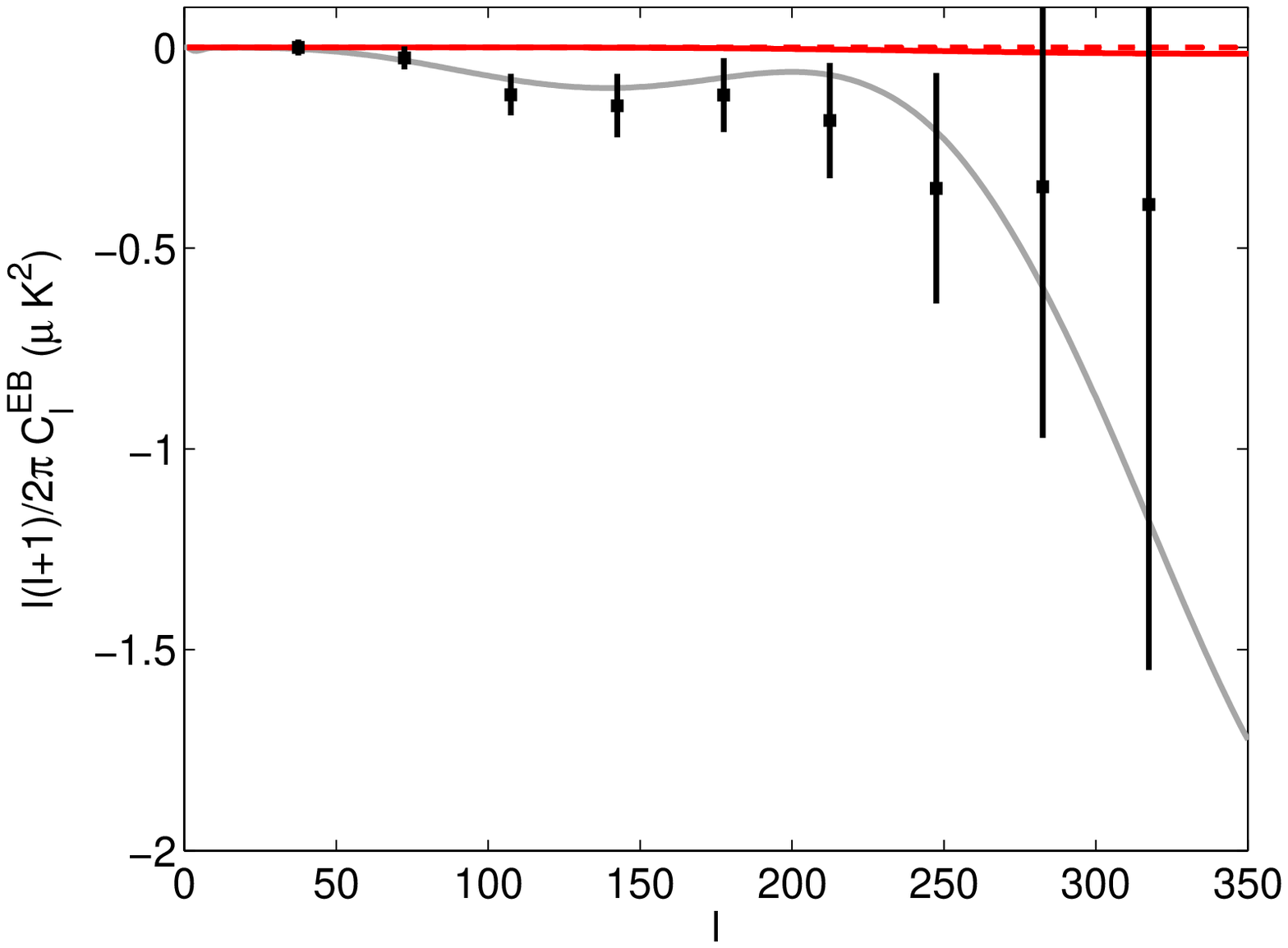}}
\caption{A plot of the effects of differential ellipticity (red solid line) and differential pointing (red dashed line) on the $TB$ (left) and $EB$ (right) power spectra. The gray line shows the power spectra with $\alpha=-2.77^{\circ}$.  The black points are the frequency-combined three-year band powers. The differential ellipticity and differential pointing are at the level of $0.2\%$ and $1.3\%$, respectively, corresponding to the upper limits reported in \cite{Takahashi2010}.  In both cases, the systematic curves correspond to ``worst-case" scenarios: the major axes of the ellipticities are separated by $45^\circ$ and differential pointing assumes a poorly-chosen scan strategy\cite{Miller2009}.}
\label{fig:tbeb_ellip}
\end{figure*}

As with differential pointing, we calculate the $TB$ and $EB$ following the construction in \cite{Shimon2008}.  As before, this is for the ``worst-case" scenario, where the major axes of the ellipticities are separated by 45$^\circ$.  From Figure \ref{fig:tbeb_ellip}, it is clear differential ellipticity can generate $TB$ power which has a different spectral shape than that produced by polarization rotation.  In addition, the $TB$ spectrum is inconsistent with the polarization rotation of $EB$.

%



\subsection{Experimental Consistency Checks}
\label{s:nulltests}

To probe the susceptibility of \bicep data to systematic effects irrespective of origin, 
\cite{Chiang2010}
created six null-tests or ``jackknife'' spectra that were used as consistency tests. These tests involve splitting 
the data in two halves and differencing them.  The two halves are chosen to illuminate systematics since signals which are common to both data sets will cancel, and the resultant jackknife will either be consistent with noise or indicate contamination.  The jackknife splits are by boresight rotation, scan direction, observing time, detector alignment, elevation coverage, and frequency, as described in \cite{Chiang2010,Barkats2013}.

\begin{figure}[b]
\includegraphics[width=0.8\columnwidth]{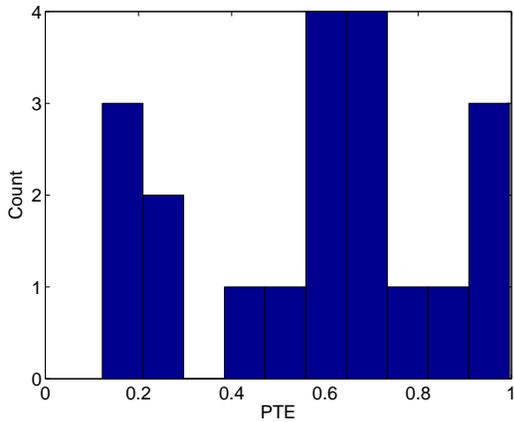}
\caption{A histogram of the probability to exceed values for the measurement of $\Delta\alpha$ in each of the jackknife spectra. There are 20 spectral combinations (which are not all independent): $TB$, $EB$, $TB+EB$, and ``all-spectra" estimators for the frequency-combined spectra for each of the five jackknife tests.}
\label{fig:PTEjacklikes}
\end{figure}

In the power spectrum analyses \cite{Chiang2010,Barkats2013}, the jackknife maps were obtained by differencing the maps for each half whereas here we ran each jackknife half through the full analysis pipeline to produce power spectra.  Unlike in the previous analyses, we did not look for consistency with zero but self-consistency between jackknife halves.  We fit rotation angles for each jackknife half and difference the resultant best-fit angles to form $\Delta\alpha$.  Only the frequency-combined spectra are used to improve the constraining power of these tests.  We then calculate the probability to exceed (PTE)  the observed $\Delta\alpha$ value by chance, 
given measurement uncertainties. These results are shown in Figure \ref{fig:PTEjacklikes}. 
If there is an instrumental systematic contribution to a detection of $\alpha$ that is tested using these null tests then excess PTE values near zero will arise.  The jackknife halves are considered consistent if they meet the following three criteria:
1) fewer than 5\% of the jackknives have PTE values smaller than 5\%, 2) none of the PTE values are excessively small (defined as $\ll$ 1\%), and 3) the PTE value from all jackknives are consistent with a uniform distribution between zero and one.
 Given the consistency of the jackknife PTEs with these criteria, systematics probed by these jackknifes are not the source of the observed polarization rotation angle.






\section{Constraints on Frequency Dependent Cosmological Birefringence} \label{s:freq_dep}

This paper has focused on the assumption that polarization rotation is independent of electromagnetic frequency. 
 However, several models feature polarization rotation that predicts a manifestly 
frequency-dependent rotation angle.  One such birefringence model has been proposed by Contaldi, Dowker, and Philpott in \cite{Contaldi2010}, hereafter called the ``CDP'' model.

Another effect which could cause frequency dependent polarization rotation would be Faraday rotation of CMB polarization due to the Milky Way's magnetic field.

\subsection{Contaldi Dowker Philpott Model}


In the CDP model, there are two electromagnetic frequency-dependent parameters ($\mu$ and $\chi$) 
leading to the following power spectra
\begin{eqnarray}
C_{\ell}^{'TT} &=& C_{\ell}^{TT} \nonumber \\
C_{\ell}^{'TB} &=& e^{-\mu} C_{\ell}^{TE} \sin(2\chi) \nonumber \\
C_{\ell}^{'EB} &=& \frac{1}{2} e^{-2\mu} \left( C_{\ell}^{EE} - 
C_{\ell}^{BB} \right) \sin(4\chi) \nonumber \\
C_{\ell}^{'TE} &=& e^{-\mu} C_{\ell}^{TE} \cos(2\chi) \nonumber \\
C_{\ell}^{'EE} &=& e^{-2\mu} C_{\ell}^{EE} \cos^2 (2\chi) + 
e^{-2\mu} C_{\ell}^{BB} \sin^2 (2\chi) \nonumber \\
C_{\ell}^{'BB} &=& e^{-2\mu} C_{\ell}^{EE} \sin^2 (2\chi) + 
e^{-2\mu} C_{\ell}^{BB} \cos^2 (2\chi)
\label{eq:CBeqnsCDP}
\end{eqnarray}
where $\mu/\chi \sim 1/ \nu$, and $\nu$ is the electromagnetic frequency 
(\emph{i.e.} 100 and 150 GHz). 
Here, $\chi$ is a frequency-dependent rotation angle, 
and $\mu$ characterizes the frequency-dependent damping parameter. The frequency-independent 
spectra are obtained in the limit $\mu\rightarrow 0$, with $\chi$ identified with $\alpha$. 

As is evident from Eq. (\ref{eq:CBeqnsCDP}), to constrain the frequency-dependent CDP model, 
the $TE$, $EE$, and $BB$ spectra must also be included in the analysis in order to break the degeneracy between 
$\chi$ and $\mu$.

\begin{figure}[h]
\includegraphics[width=0.8\columnwidth]{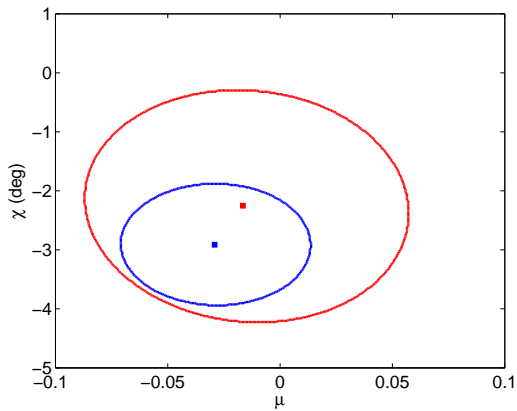}
\caption{Best fit $\mu$ and $\chi$ values for the CDP model for 100 GHz (red point) and 150 GHz (blue point) from the all-spectra estimator, along with their 68\% confidence interval contours (red and blue contours respectively).}
\label{fig:freq_dep}
\end{figure}

   
 The results for the damping parameter and rotation angle $\chi$ are presented in Table \ref{table:BicepCBCDP} and Figure \ref{fig:freq_dep}. The inferred $\mu$ is consistent with zero and $\mu/\chi$ is not inversely proportional to $\nu$, thus there is no compelling evidence for frequency dependent birefringence in the CDP picture.

\begin{table}[h]
\caption{Maximum likelihood values for the damping parameter, $\mu$, and the rotation angle, $\chi$, along with their 1$\sigma$ error bars for the CDP model.}

\begin{tabular}{|c|c|c|}
\hline Frequency (GHz) & $\mu$ & $\chi$ (degrees)\\
\hline $100$ & $-0.017^{+0.073}_{-0.076}$ & $-2.25^{+2.02}_{-2.02}$\\
\hline $150$ & $-0.029^{+0.042}_{-0.043}$ & $-2.91^{+1.02}_{-1.02}$\\
\hline
\end{tabular}

\label{table:BicepCBCDP}
\end{table}

\subsection{Faraday Rotation of Galactic Magnetic Field}

Faraday rotation due to the Milky Way's magnetic field predicts frequency-dependent polarization rotation proportional to the inverse of the frequency squared \cite{FR2013}.  Scaling the 150 GHz ``all-spectra" estimator $\alpha$ by (100/150)$^{-2}$, we would expect to see polarization rotation of the 100 GHz spectra consistent with an $\alpha = -6.55^{+2.34}_{-2.39}$ degrees, whereas our 100 GHz ``all-spectra" estimator results in a polarization rotation of $\alpha = -2.27^{+2.06}_{-2.02}$, corresponding to a $1.37\sigma$ discrepancy.  While there is some tension between the predicted $\alpha$ and the measured $\alpha$, Faraday rotation cannot be ruled out as the cause of the rotation.


\section{Self-Calibrated Upper Limit on Tensor to Scalar Ratio} \label{s:self_cal}

If the polarization rotation is systematic in nature, the derived rotation angle can be used to calibrate the detector polarization orientations \cite{KSY}.  The three-year ``all-spectra" rotation angles were added to the polarization orientations treating the frequency bands as independent, i.e., only the 100 GHz (150 GHz) derived rotation angle was added to the 100 GHz (150 GHz) detectors.  These ``self-calibrated" polarization orientation angles were propagated through the power spectrum analysis pipeline \cite{Barkats2013}.  The self-calibrated power spectra were analyzed for residual polarization rotation which yielded a rotation angle $\alpha = +0.01^\circ \pm 0.86^\circ$ from the frequency-combined all-spectra estimator, consistent with zero, as expected.

Any polarization rotation, regardless of cosmic or systematic origin, will positively bias $r$ since $E$-mode power will be leaked into the $B$-mode spectrum (Equation \ref{eq:CBeqnsXia}).  There is also a reduction to the $B$-mode power spectrum
due to $B$-modes leaking to $E$-modes, however since the $E$-modes are significantly larger than the $B$-modes, the net result is a positive bias on the $B$-mode power spectrum.  From the self-calibrated three-year power spectra, following the procedure in \cite{Barkats2013}, we find the upper limit on the tensor-to-scalar ratio reduces from $r<0.70$ to $r<0.65$ at 95\% confidence.  From simulations, we find that the bias on $r$ from self-calibration with no underlying polarization rotation is less that 0.01.


\section{Conclusion} \label{s:conclusion}

The \bicep three-year data, when analyzed using detector polarization orientations
from our standard dielectric sheet calibrator, show non-vanishing $TB$ and $EB$ spectra 
consistent with an overall polarization rotation of $-2.77^\circ \pm 0.86^\circ$ at 3.22$\sigma$ significance.
The significance for non-zero rotation of astrophysical origin is only $1.78\sigma$, given
the $1.3^\circ$ systematic uncertainty on our orientation calibration which adds in quadrature.  
This result passes experimental consistency tests which probe for systematic differences of
polarization rotation in various subsets of data.  
We rule out beam systematics as significant, and identify polarization orientation miscalibration
as the primary concern among instrumental systematics.  
Isotropic cosmic birefringence can not be excluded, though is is degenerate with a polarization miscalibration.  
The data show no compelling evidence for frequency-dependent isotropic cosmic birefringence models.  
An alternate use of the measurements described here is to self-calibration the detector polarization orientations, 
at the expense of losing constraining power on isotropic cosmological birefringence \cite{KSY}.  
Self-calibrating the \bicep three-year data reduces the upper limit on the tensor-to-scalar 
ratio from $r<0.70$ to $r< 0.65$ at 95\% confidence.  

Future CMB polarimeters with improved polarization calibration methods will be needed to 
break the degeneracy between polarization rotation and detector polarization orientation uncertainty.  
In addition to the CMB, complimentary astronomical probes such as the polarization orientation of 
radio galaxies and quasars \cite{RGCB2010,Kamionkowski2010} can help constrain cosmological birefringence.  
However, these objects can only constrain cosmic birefringence over a limited range of redshifts and 
only along particular lines-of-sight, whereas CMB polarization can be used to constrain cosmic 
birefringence over the entire sky and is sensitive to effects accrued over the history of the entire Universe.  
Polarization angles calibrated with current man-made or astronomical sources are accurate enough for 
current generation $B$-mode measurements, but are insufficiently characterized for cosmic birefringence searches.  
Based on \bicep\ experiences with systematic uncertainties on polarization orientation calibration
reported in this paper, improved far-field calibrators have been developed for BICEP2 and
other future experiments.
The revolutionary discovery potential of a detection of cosmic birefringence motivates the development of 
more accurate hardware calibrators and further investigation of astronomical sources, 
to achieve a precision of $\ll 0.5^\circ$.  Ultimately, a combination of precisely 
understood man-made and astronomical sources will allow for powerful constraints on parity 
violation which will come concomitantly with bounds on the physics of inflation.

%
%

\begin{acknowledgments}

\bicep was supported by NSF Grant No. OPP-0230438, Caltech PresidentÕs Discovery Fund, Caltech PresidentÕs Fund PF-471, JPL Research and Technology Development Fund, and the late J. Robinson. This analysis was supported in part by NSF CAREER award No. AST-1255358 and the Harvard College Observatory, and J.M.K. acknowledges support from an Alfred P. Sloan Research Fellowship. B.G.K acknowledges support from NSF PECASE Award No. AST- 0548262.  N.J.M.'s research was supported by an appointment to the NASA Postdoctoral Program at Goddard Space Flight Center, administered by Oak Ridge Associated Universities throughÊa contract with NASA.  M.S. acknowledges support from a grant from Joan and Irwin Jacobs. We thank the South Pole Station staff for helping make our observing seasons a success. We also thank our colleagues in the ACBAR, BOOMERANG, QUAD, BOLOCAM, SPT, WMAP and Planck experiments, as well as Kim Griest, Amit Yadav, and Casey Conger for advice and helpful discussions, and Kathy Deniston and Irene Coyle for logistical and administrative support. We thank Patrick Shopbell for computational support at Caltech and the FAS Science Division Research Computing Group at Harvard University for providing support to run all the computations for this paper on the Odyssey cluster.

\end{acknowledgments}

\bibliography{cb_references}

\end{document}